\documentstyle[prl,twocolumn,aps,epsf]{revtex}

\def\nbZ{{\mathchoice {\hbox{$\sf\textstyle Z\kern-0.4em Z$}}
{\hbox{$\sf\textstyle Z\kern-0.4em Z$}}
{\hbox{$\sf\scriptstyle Z\kern-0.3em Z$}}
{\hbox{$\sf\scriptscriptstyle Z\kern-0.2em Z$}}}}

\begin{document}

\twocolumn[\hsize\textwidth\columnwidth\hsize\csname @twocolumnfalse\endcsname
\draft
\tolerance 500

\title{Pairing of Cooper Pairs in a Fully Frustrated Josephson Junction Chain}
\author{Benoit Dou\c{c}ot$^{1,2}$ and Julien Vidal$^{3}$,  }

\address{$^1$ Laboratoire de Physique de la Mati\`ere Condens\'ee, 
CNRS UMR 8551,
\'Ecole Normale Sup\'erieure,\\
24, rue Lhomond, 75231 Paris Cedex 05 France}

\address{$^2$ Laboratoire de Physique Th\'{e}orique et
Hautes \'Energies, CNRS UMR 7589, Universit\'{e}s Paris 6 et 7,\\
4, place Jussieu, 75252 Paris Cedex 05 France}

\address{$^3$ Groupe de Physique des Solides, CNRS UMR 7588,
Universit\'{e}s Paris 6 et 7,\\
2, place Jussieu, 75251 Paris Cedex 05 France}

\maketitle

\begin{abstract}
We study a one-dimensional Josephson junction chain embedded in a 
magnetic field. We show that when
the magnetic flux per elementary loop equals half the superconducting 
flux quantum $\phi_0=h/2e$, a local
$\nbZ_2$ symmetry arises. This symmetry is responsible for a nematic 
Luttinger liquid state associated to
bound states of Cooper pairs. We analyze the phase diagram and we 
discuss some experimental possibilities
to observe this exotic phase.
 
\end{abstract}


\vskip2pc]

%
%
%%%%%%%%%%%%%%%%%%%%%%%
%Introduction
%%%%%%%%%%%%%%%%%%%%%%%
%
%
During the last twenty years, Josephson junction arrays have proved 
to be very good tools to investigate
classical and quantum phase transitions \cite{Fazio_Review}.
Recently, much attention has been payed to systems which display 
highly degenerate classical ground states
\cite{Korshunov_T3} due to the presence of Aharonov-Bohm 
cages\cite{Vidal_Cages}.
Interestingly, a glassy vortex phase without disorder has been 
predicted for such two-dimensional ($2D$)
structures \cite{Cataudella_T3} in agreement with experimental observations
\cite{Pannetier_T3}.  In the present work, we investigate the 
influence of quantum
fluctuations on such systems in a $1D$ model introduced in 
\cite{Vidal_Chapelet}.
In this remarkably simple example, the huge classical degeneracy is a 
direct consequence of a local
$\nbZ_2$ symmetry which is unbroken in the presence of quantum 
fluctuations. We show that this may
stabilize an unusual nematic Luttiger liquid (LL) phase in which 
charge $4e$ bound states of Cooper pairs
are the elementary objects.
%%%%%%%%%%%%%%%%%%%%%%%%%%%%%%%%%%%%%%%%%%%%%%%%%%%%%%%%%%%%%%%%%%%%%
%%%%%%%%%%%%%%%%%%%%%%%%%%%%%%%%%%%%%%%%%%%%%%%%%%%%%%%%%%%%%%%%%%%%%

\begin{figure}
\centerline{\epsfxsize=100mm
\hspace{8mm}\epsffile{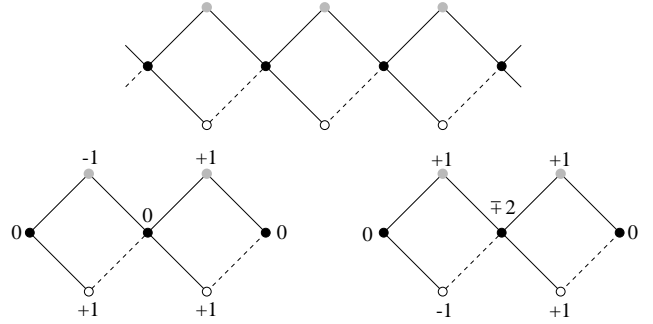}}
\vspace{-2mm}
\caption{The chain of loops and  the three  (nonnormalized) cage
eigenstates corresponding to $\varepsilon_0$ (left) and 
$\varepsilon_\pm$ (right). The
dashed line symbolizes the hopping term
$-t_{\rm J} \, {\rm e}^{-{\rm i} \gamma}$.}
\label{fig:chain}
\end{figure}

We consider the chain of loops shown in Fig.~\ref{fig:chain} embedded 
in a uniform magnetic
field. We denote by $\phi$ the magnetic flux per elementary plaquette 
and we set $\gamma=2\pi \phi/ \phi_0$
where $\phi_0=h/2e$ is the superconducting flux quantum.
Each site of this lattice is supposed to be occupied by a 
superconducting island.
A convenient description of the low-energy Hilbert space of this 
system involves local boson operators
$a^{\dagger}_n, b^{\dagger}_n, c^{\dagger}_n$  $(a_n, b_n, c_n)$ that 
create (destroy) Cooper
pairs on the three types of islands of the lattice, respectively 
represented by black, grey and
white circles in Fig.~\ref{fig:chain}.
The system is described by the following Josephson coupling Hamiltonian~:
%
%
%%%%%%%%%%%%%%%%%
\begin{equation}
H_{\rm J}= -t_{\rm J} \sum_n a^{\dagger}_n (b_n + c_n + b_{n-1} + 
{\rm e}^{-{\rm i} \gamma} c_{n-1}) +
\mbox{h.\,c.}
\label{HJ1}
\end{equation}
%%%%%%%%%%%%%%%%%
%
%
We first focus on the special value $\gamma=\pi$ (half a flux quantum 
per loop). As shown
in \cite{Vidal_Chapelet}, this Hamiltonian has, in this case,  a 
single particle spectrum composed of three
highly degenerate flat bands
$\varepsilon_0=0$, $\varepsilon_\pm=\pm 2 t_{\rm J}$. The 
corresponding eigenstates can be chosen as
strictly localized (cage states) around each fourfold coordinated 
site (see Fig.~\ref{fig:chain}). This
leads naturally to the notion of Aharonov-Bohm cages discussed in 
\cite{Vidal_Cages}.
%
%
%%%%%%%%%%%%%%%%%%%%%%%%%%%%%%%%%
%%%%%%%%%%%%%%%%%%%%%%%%%%%%%%%%%
%
%
Let us introduce the set of boson operators $A^{\dagger}_{\alpha,n}$ 
($A_{\alpha,n}$) that
create (destroy) one Cooper pair with energy $\varepsilon_\alpha$ 
($\alpha=0,\pm$) in a cage state
localized around the $n$th fourfold site. These operators can be 
simply expressed as a linear combination
of the operators
$a^{\dagger}_n, 
b^{\dagger}_n,c^{\dagger}_n,b^{\dagger}_{n-1},c^{\dagger}_{n-1}$ 
only, whose coefficients
are given in Fig.~\ref{fig:chain} so that we get~:
%
%
%%%%%%%%%%%%%%%%%
\begin{equation}
H_{\rm J}= \sum_{n,\alpha} \varepsilon_\alpha A^{\dagger}_{\alpha,n} 
A_{\alpha,n}
\mbox{ . }
\label{HJ2}
\end{equation}
%%%%%%%%%%%%%%%%%
%
%

In this present form, $H_{\rm J}$ clearly exhibits a local $U(1)$ symmetry.
We shall now study the effect of boson-boson interaction on this 
symmetry. Therefore,
we consider a real valued function $n \mapsto s_n$, and we construct 
a unitary operator
$U_s$ defined by~: $ U_s A_{\alpha,n} U^{-1}_s= {\rm e}^{-{\rm i} 
s_n} A_{\alpha,n}$
which commutes with $H_{\rm J}$. Using the precise form of the cage 
states, we easily obtain~:
%
%
%%%%%%%%%%%%%%%%%
\begin{eqnarray}
U_s a^{\dagger}_{n} a_{n}  U^{-1}_s &=& a^{\dagger}_{n} a_{n}
\label{defU1}\\
U_s b^{\dagger}_{n} b_{n}U^{-1}_s &=&\cos^2(\Delta_n)  b^{\dagger}_{n} b_{n} +
\sin^2(\Delta_n)  c^{\dagger}_{n} c_{n} + z_n\label{defU2}\\
U_s c^{\dagger}_{n} c_{n}U^{-1}_s &=&\sin^2(\Delta_n)  b^{\dagger}_{n} b_{n} +
\cos^2(\Delta_n)  c^{\dagger}_{n} c_{n} - z_n \label{defU3}
\mbox{ , }
\end{eqnarray}
%%%%%%%%%%%%%%%%%
%
%
where we have set $z_n={\rm i} \sin(2\Delta_n) (b^{\dagger}_{n} 
c_{n}-c^{\dagger}_{n} b_{n})/2$ and
$\Delta_n=(s_{n+1}-s_n)/2$. From these transformation laws, we can 
readily see that any interaction term
involving the bilinear operators $a^{\dagger}_{m} a_{m}$, 
$b^{\dagger}_n b_n+c^{\dagger}_n c_n$ preserves
the local $U(1)$ symmetry. Physically, this symmetry implies that the 
total number of bosons in each cage
is sepa\-rately conserved and the system remains an insulator. 
However, this symmetry is
fragile since it is easy to find two-body interactions which break 
it. For instance, a Hubbard-like
interaction term $\sum_n (a^{\dagger}_{n} a_{n})^2 + 
(b^{\dagger}_{n} b_{n})^2 + (c^{\dagger}_{n}
c_{n})^2$, has this effect which is manifested by the appearance of 
delocalized two-particle bound states
discussed in \cite{Vidal_Chapelet}. An exciting feature of this 
system is that this type of interaction
still preserves a subgroup of the full $U(1)$ corresponding to a 
local $\nbZ_2$ symmetry. This subgroup
corresponds to $s_n=0 \,[\pi]$ for all $n$. With
this restriction, it is easy to check that the operator
$b^{\dagger}_m b_m b^{\dagger}_n b_n+c^{\dagger}_m c_m c^{\dagger}_n 
c_n$ commutes with $U_s$ for all
$(m,n)$. This local $\nbZ_2$ symmetry has an important physical 
consequence since it means that the parity
of the total number of bosons in each cage is separately conserved. 
Therefore, if two-particle
interactions lead to coherent transport through the chain for a 
many-boson system,  quasi off-diagonal
long range order may occur only for composite objects built with an 
even number of original bosons {\it
i.~e.}, here, of Cooper pairs. In other words, a superconducting 
Josephson junction chain with this
geometry and half a flux quantum per loop may realize a quasi-Bose 
condensate (in fact a LL) of charge $4e$
composite bosons~!

To discuss in more details the physics of this system, it is useful to rephrase these
symmetry considerations in the language of quantum rotor mo 
dels. These offer the advantage of an
intuitively simple classical limit defined from the phase of 
superconducting order parameter. Formally,
we introduce three phase fields $\theta_n, \varphi_n, \chi_n$ and 
their canonically conjugate fields
$\Pi_{\theta,n}, \Pi_{\varphi,n}, \Pi_{\chi,n}$, which are related to 
the local Bose operators by~:
$a^{\dagger}_n=\Pi^{1/2}_{\theta,n} \, {\rm e}^{{\rm i} \theta_n}$,
$b^{\dagger}_n=\Pi^{1/2}_{\varphi,n} \, {\rm e}^{{\rm i} \varphi_n}$,
$c^{\dagger}_n=\Pi^{1/2}_{\chi,n} \, {\rm e}^{{\rm i} \chi_n}$.
Assuming that the local particle number fluctuations are small, we 
get the  quantum phase
Hamiltonian \cite{Fazio_Review}
%
%
%%%%%%%%%%%%%%%%%
\begin{eqnarray}
H&=& {E_{\rm C}\over 2} \sum_{n}
  \Pi^2_{\theta,n}+ \Pi^2_{\varphi,n}+\Pi^2_{\chi,n}  - \nonumber \\
&&E_{\rm J} \sum_{n}  \cos(\theta_n-\varphi_n)+ 
\cos(\theta_n-\chi_n)+ \nonumber\\
&&\cos(\theta_n-\varphi_{n-1})+\cos(\theta_n-\chi_{n-1}-\gamma)
\mbox{ , }
\label{H}
\end{eqnarray}
%%%%%%%%%%%%%%%%%
%
%
where $E_{\rm C}$ is the charging energy and $E_{\rm J}$ the 
Josephson coupling between islands.
Note that the present modelling of capacitive effects is not meant to be
very realistic, since for the sake of simplicity, we have not taken 
into account off-diagonal elements of
the capacitance matrix. This choice corresponds to a local 
Hubbard-like interaction term between Cooper
pairs. For convenience, we set $\sqrt{E_{\rm C} E_{\rm J}}=1$.

\begin{figure}
\centerline{\epsfxsize=75mm
\epsffile{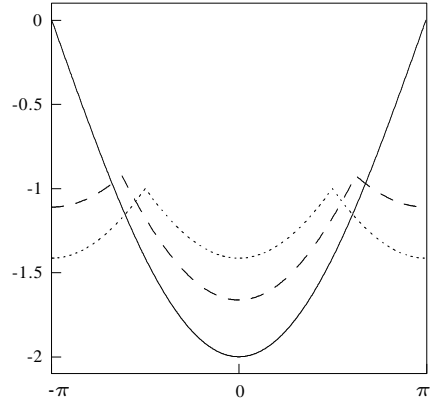}}
\vspace{-13mm}
\caption{Behaviour of $F(x_n)$ for $\gamma=0$ (solid line), 
$\gamma=3\pi/4$ (dashed line) and
$\gamma=\pi$ (dotted line). }
\label{fig:F}
\end{figure}

\begin{figure}
\centerline{\epsfxsize=78mm
\epsffile{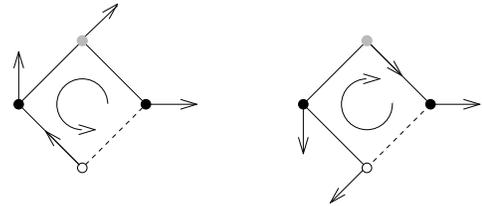}}
\caption{Two possible classical ground states of $H$ with different chirality.}
\label{fig:Z2}
\end{figure}

The classical ground state of $H$ is easily obtained for any 
$\gamma$. Indeed, if we
set $x_n=\theta_{n+1}-\theta_{n}-\gamma/2$, and eliminate $\varphi_n$ 
and $\chi_n$, minimizing $H$
is equivalent to minimize
$F(x_n)=-\left|\cos\left(x_n/2+\gamma/4\right)\right|-\left|\cos\left( 
x_n/2-\gamma/4\right)\right|$  for
all $n$. As shown in Fig.~\ref{fig:F}, $F$ has two local minima in 
$x_n=0$ and $x_n=\pi$. For $0<
\gamma < \pi$, one has  $F(0) < F(\pi)$ and the classical ground 
state is unique (up to a global $U(1)$
degeneracy).  By contrast, for $\gamma=\pi$, one has $F(0)=F(\pi)$ so 
that, for a given plaquette, we
obtain two degenerate ground states (up to a global translation of 
the phase variables) which are
illustrated in Fig.~\ref{fig:Z2}.
%
%
%%%%%%%%%%%%%%%%%%%%%%%%%%%%%%%%%
%%%%%%%%%%%%%%%%%%%%%%%%%%%%%%%%%
%
%

\noindent These states only differ in the sign of the
superconducting currents which circulate around the plaquette. For a 
chain made up of $N$ loops, we thus
get $2^N$ degenerate classical ground states up to a global 
translation of the phase variables.
%
%
%%%%%%%%%%%%%%%%%%%%%%%%%%%%%%%%%
%%%%%%%%%%%%%%%%%%%%%%%%%%%%%%%%%
%
%

This huge degeneracy is a direct consequence of a local $\nbZ_2$ 
symmetry of $H$.  Note that $H$ is not
invariant under the full local $U(1)$ group related to the 
Aharonov-Bohm cages. This occurs since the
Josephson term in $H$ may be written as a strongly non linear 
expression of the basic local Bose
operators. For the $\nbZ_2$  transformations, it is an easy task to 
translate the $U_s$ operators in
the language of phase variables.  An interesting local $\nbZ_2$ 
transformation is provided by a kink in
the $s_m$'s. Let $U_n$ be the transformation defined by $s_m=0$ for 
$m \leq n$ and $s_m=\pi$ for $m > n$.
This transformation does not modify the phase variables for $m\leq n$ 
whereas it shifts them by
$\pi$ if $m>n$ and it permutes $\varphi_n$ and $\chi_n$. Thus, its 
main physical effect is to
change the currents flowing around the plaquette located between $n$ 
and $n+1$ into their
opposite value. From this description, we deduce that starting from a 
given classical ground state, we
may generate any other ground state by applying a finite sequence of 
such $U_n$ operators. We also see
that, in the classical limit considered here, the local $\nbZ_2$ 
symmetry is spontaneously broken,
yielding ground states with well-defined local circulating supercurrents.
Note that $U_n$ also leaves  most conjugate variables unchanged
except $\Pi_{\varphi,n}$ and $\Pi_{\chi,n}$ which are exchanged. As a 
result, we may add to $H$ any term
involving these conjugate fields without breaking the local $\nbZ_2$ 
symmetry, provided the spatial
symmetry between the $\varphi_n$ and $\chi_n$ degrees of freedom is respected.
Experimentally, this would require a tight control of offset charges
since these may seriously alter an otherwise excellent geometrical 
symmetry of the chain.

For real systems, it may become important to take into account 
quantum fluctuations of the phase
variables, especially when the superconducting islands are so small 
that their charging energy
$E_{\rm C}$ can no longer be neglected in comparison to the Josephson 
coupling energy $E_{\rm J}$. For a
single loop and $\gamma=\pi$, these quantum fluctuations have been shown,
theore\-tically \cite{Caldeira_Leggett} and experimentally 
\cite{Friedman,Van_der_Wal}, to induce tunneling
between the two degenerate classical ground states shown in 
Fig.~\ref{fig:Z2}. The true quantum mechanical
ground state is therefore a macroscopic linear superposition of these 
two classical states and provides a
simple  example of a ``Schr{\"o}dinger cat".  For a system with $N$ 
loops, eigenstates are classified
according to the various irreducible representations of the local 
$\nbZ_2$ group which mix all the
$2^N$ classical ground states.
One of our next goals is to describe how quantum fluctuations lift 
the degeneracy among these
representations, which is an artifact of the classical limit.

At small $g=\sqrt{E_{\rm C}/E_{\rm J}}$, the properties of the system 
are actually very similar to those
of a quantum $XY$ model.  For small $\gamma$, we thus expect the 
infinite chain to be in a LL phase for
$g < g^*(\gamma)$ and in a gapful insulating (I) phase for $g > g^*(\gamma)$.
The transition at $g^*(\gamma)$ is of Berezinskii-Kosterlitz-Thouless (BKT)
type \cite{BKT}. Simple spin wave calculations using the harmonic
approximation of $H$ around its classical ground state predict
$g^*(\gamma) = \sqrt{\cos(\gamma/4)}\, g^*(0)$ with $g^*(0)=\pi 
\sqrt{3}/2$. The main effect of the
magnetic field, in this simple approximation, is thus to replace $g$ 
by an effective
$g_{eff}=g/\sqrt{\cos(\gamma/4)}$ which controls all the correlation 
function exponents in the LL phase.

To analyze the effect of the quantum fluctuations in the vicinity of 
$\gamma=\pi$ where the additional
local $\nbZ_2$ symmetry emerges, it is convenient to eliminate the 
twofold coordinated islands to get a
simple description of the low-energy physics of this system. 
Therefore, instead of $H$, we now consider
the following Hamiltonian~:
%
%
%%%%%%%%%%%%%%%%%
\begin{eqnarray}
H_{XY} &=&  \sum_{n}
{g' \over 2} \Pi^2_{\theta,n}-
{1 \over g'} \{ \cos[p(\theta_n-\theta_{n+1}-\gamma / 2)]+\nonumber\\
&&\epsilon \cos(\theta_n-\theta_{n+1}- \gamma/ 2)\}
\mbox{ . }
\label{HXY}
\end{eqnarray}
%%%%%%%%%%%%%%%%%
%
%
The parameter $g'$ is provided by fitting the exponent of the 
correlation function
$\langle {\rm e}^{{\rm i} (\theta_m-\theta_n)} \rangle$ in the 
semi-classical regime with its value
obtained with $H$ in the harmonic approximation. This choice leads to 
$g'=2^{5/4} 3^{-1/2} g$.
The parameter $\epsilon=4|\gamma-\pi|$ is  determined from the energy 
splitting between the two local
minima of the single loop potential energy. Finally, in our case, we 
have $p=2$ but we discuss
thereafter the properties of $H_{XY}$ for an arbitrary $p$.

The Hamiltonian $H_{XY}$ has a local $\nbZ_p$ symmetry at 
$\epsilon=0$, corresponding to the local
transformations $T_a$~: $\theta_j \mapsto \theta_j+2 \pi a_j / p$ 
where $a_j$ is an integer. The
irreducible representations of this group are easily obtained in a 
basis which diagonalizes
simultaneously the $\Pi_{\theta,j}$'s. For a state $|\psi\rangle$ such that
$\Pi_{\theta,j} |\psi\rangle= l_j |\psi\rangle$, where $l_j$ is an 
integer, we have
$T_a|\psi\rangle=\exp({\rm i} \sum_j {2 \pi \over p} \, l_j \,a_j ) 
|\psi\rangle$.
Writing $l_j= m_j+p \,n_j$ with $m_j$ and $n_j$ integers and $0 \leq 
m_j \leq p-1$, we find that the set
of $m_j$'s completely specifies the irreducible representation of the 
local $\nbZ_p$ group. For each such
representation, the action of the corresponding projector on the 
approximate gaussian ground state of
$H_{XY}$ produces a natural trial wave-function at least when $g' \ll 1$.
We have computed the expectation value of $H_{XY}$ on these states.
Doing so, we noticed that $2\pi/p$-tunnel processes occuring on 
different lattice sites are mostly
uncorrelated. Neglecting completely these correlations we get~:
%
%
%%%%%%%%%%%%%%%%%
\begin{equation}
\langle H_{XY} \rangle={g' L^2 \over 2N}-{C {\rm e}^{-f} \over
g'} \sum_{j} \cos \left[{2 \pi \over p}(m_j- L/ N)\right]
\mbox{,}
\label{Expectation2}
\end{equation}
%%%%%%%%%%%%%%%%%
%
%
up to a constant energy independent of the representation and  to 
factors of order ${\rm e}^{-2 f}$.
In (\ref{Expectation2}), $L=\sum_j l_j$ is the total angular 
momentum, $C$ is a number close to $2\pi^2-8$
at small $g'$, and $f\simeq 4 \pi / p g'$. The ground state is 
therefore obtained by choosing the identity
representation of the local $\nbZ_p$ group ($m_j=0$).

Next, we see that the term proportionnal to $\epsilon$ couples 
different irreducible representations of the
local $\nbZ_p$ group. When $p=2$ the action of this perturbation on 
the $2^N$ low-energy trial states
just discussed, is well described by a quantum Ising model in a 
transverse magnetic field~:
%
%
%%%%%%%%%%%%%%%%%
\begin{eqnarray}
H_{\rm I} &=& -{1\over g'} \left( C {\rm e}^{-f} \sum_n \sigma_n^X +
{D \epsilon } \sum_n \sigma_n^Z \sigma_{n+1}^Z \right)
\mbox{ , }
\label{HIsing}
\end{eqnarray}
%%%%%%%%%%%%%%%%%
%
%
where $D$ is close to $1$ for small $g'$. In terms of these Ising 
variables, the local $\nbZ_2$ symmetry
corresponds to interchanging the
$|+\rangle$ and $|-\rangle$ states on any given subset  of sites. It 
is therefore implemented by
the
$\sigma_n^X$ operators.
This model has a continuous phase transition (in the universality 
class of the $2D$
Ising model) at its self-dual point which corresponds to 
$\epsilon=(C/D) \, {\rm e}^{-2 \pi/ g'}$.
Furthermore, it is easy to see that $H_{XY}$ exhibits a BKT 
transition at $\gamma=\pi$ (so $\epsilon=0$)
corresponding to the loss of quasi long range order for the nematic 
order parameter ${\rm e}^{2 {\rm i}
\theta}$. In the harmonic approximation, this occurs at $\tilde 
g(\pi)=g^*(\pi)/4$.

\begin{figure}
\centerline{\epsfxsize=75mm
\epsffile{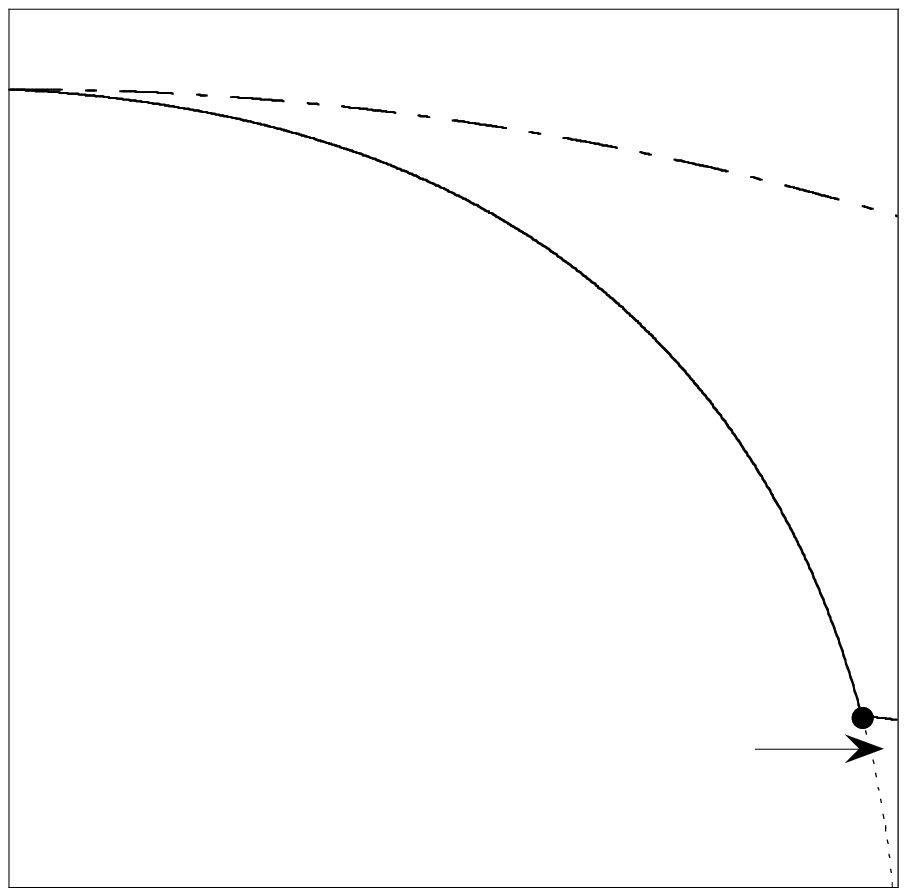}}
{\small
\vspace{1pt}
\vspace{-70pt}
\hspace{185pt}
$\pi$

\hspace{120pt}
$\gamma$

\vspace{-157pt}
\hspace{24pt}
$g^*(0)$

\vspace{8pt}
\hspace{195pt}
$g^*(\pi)$

\vspace{70pt}
\hspace{195pt}
$\tilde g(\pi)$

\vspace{18pt}
\hspace{40pt}
$0$

\vspace{-3pt}
\hspace{50pt}
$0$

\vspace{-41pt}
\hspace{90pt}
Nematic LL ($4e$)

\vspace{-61pt}
\hspace{90pt}
  LL ($2e$)

\vspace{-38pt}
\hspace{148pt}
Insulator

\vspace{60pt}
\hspace{174pt}
$Q$

}
\vspace{50pt}
\caption{Schematic phase diagram as a function of
$g=\sqrt{E_{\rm C}/E_{\rm J}}$,
and $\gamma=2\pi \phi/\phi_0$. LL stands for Luttinger liquid.}
\label{fig:Phase_Diagram}
\end{figure}

Gathering these informations obtained in various limits, we get the 
phase diagram drawn in
Fig.~\ref{fig:Phase_Diagram}. Besides the two familiar phases, namely 
the (I) phase at large
$g$ and the LL phase characterized by an algebraic order of the ${\rm 
e}^{{\rm i}
\theta}$ order parameter, the most interesting result is the presence 
of a remarkable nematic Luttinger
liquid (NLL) phase which may be viewed as a quasi-ordered condensate 
of pairs of Cooper pairs associated
to the order parameter ${\rm e}^{2 {\rm i} \theta}$.
%
%
%%%%%%%%%%%%%%%%%%%%%%%%%%%%%%%%%
%%%%%%%%%%%%%%%%%%%%%%%%%%%%%%%%%
%
%
The dotted line of $2D$ Ising-type separates the two Luttinger 
phases. The Ising order parameter
is vanishing in the NLL phase and builds up in the conventional LL 
phase where its square is
proportional to the algebraically decaying part of the ${\rm e}^{{\rm 
i} \theta}$ autocorrelation function.
The physical picture presented here has already been uncovered by Lee 
and Grinstein \cite{Lee_Grinstein}
in the framework of a $2D$ classical $XY$ model with squared cosine 
interaction. In their language, the
BKT transition from the $2e$ LL phase to the I phase involves $2\pi$ 
vortices of the $\theta$
field, by contrast to $\pi$ vortices between the NLL phase and the I 
phase. The nature of the
multicritical point $Q$ remains mysterious to us.
Note that the coupling between $XY$ and Ising-like degrees of freedom 
has been  extensively
studied
\cite{Teitel_XY,Halsey_XY,Korshunov_XY,Granato_86,Granato_91,Li_Ciepak},
already in the context of
fully frustrated  Josephson juntion arrays. Nevertheless, in these 
works, the Ising order parameter has a
very different nature since it describes the possible melting of the 
vortex lattice. It implies that the
$XY$ transition has to appear at lower temperatures than the Ising
transition \cite{Teitel_XY,Halsey_XY,Korshunov_XY}. In our case and 
in the vicinity of $\gamma=\pi$, the
$XY$ transition occurs for larger $g$ than the Ising one.  So, the 
effect of the Ising domain walls on the
phase stiffness is very different in the two situations.

To conclude, we have established the possibility of macroscopic 
condensation (in the sense of a LL) of
charge $4e$ objects in a Josephson junction chain for $\gamma\simeq \pi$.
Experimentally, it may be possible to detect this binding of Cooper 
pairs by connecting such a network to
superconducting leads. Indeed, we expect here a phenomenon analogous 
to Andreev's reflection~: an
ordinary Cooper pair of charge $2e$ entering the chain at low energy 
(compared to the gap between
different representations of the local $\nbZ_2$ group ) will leave 
behind a pair of charge $-2e$ so that
a charge $4e$ composite object may propagate along the chain. Another 
possibility is to close the chain
into a large ring. In this geometry, we expect quantum oscillations 
of the global current with
respect to the magnetic flux across the ring with an elementary 
period $\phi_0/2$ as long as
$\gamma\simeq \pi$.

\bigskip
We thank P. Lecheminant, D. Mouhanna, B. Pannetier and E. Serret for 
fruitful and stimulating discussions.
%
%
%
%\bibliography{bibliotheque}

\begin{thebibliography}{10}

\bibitem{Fazio_Review}
R. Fazio and {H. van der Zant}, Phys. Rep. {\bf 355},  235  (2001).

\bibitem{Korshunov_T3}
S.~E. Korshunov, Phys. Rev. B {\bf 63},  134503  (2001).

\bibitem{Vidal_Cages}
J. Vidal, R. Mosseri, and B. Dou\c{c}ot, Phys. Rev. Lett. {\bf 81},  5888
(1998)~; J. Vidal, P. Butaud, B. Dou\c{c}ot, and R. Mosseri, Phys. 
Rev. B {\bf 64}, 155306  (2001).

\bibitem{Cataudella_T3}
V. Cataudella and R. Fazio, cond-mat/0112307.

\bibitem{Pannetier_T3}
{B. Pannetier {\it et~al.}}, Physica C {\bf 352},  41  (2001)
{E. Serret {\it et~al.}}, unpublished.

\bibitem{Vidal_Chapelet}
J. Vidal, B. Dou\c{c}ot, R. Mosseri, and P. Butaud, Phys. Rev. Lett. {\bf 85},
   3906  (2000).

\bibitem{Caldeira_Leggett}
A.~O. Caldeira and A.~J. Leggett, Phys. Rev. Lett. {\bf 46},  211  (1981).

\bibitem{Friedman}
{J.~R. Friedman {\it et~al.}}, Nature {\bf 406},  43  (2000).

\bibitem{Van_der_Wal}
{C. H. van der Wal {\it et~al.}}, Science {\bf 290},  773  (2000).

\bibitem{BKT}
V.~L. Berezinskii, Zh. Eksp. Teor. Fiz. {\bf 59},  907  (1970) [Sov. 
Phys. JETP {\bf 32},  493  (1971)]~;
J.~M. Kosterlitz and D.~J. Thouless, J. Phys. C {\bf 6},  1181  (1973).

\bibitem{Lee_Grinstein}
D.~H. Lee and G. Grinstein, Phys. Rev. Lett. {\bf 55},  541  (1985).

\bibitem{Teitel_XY}
S. Teitel and C. Jayaprakash, Phys. Rev. B {\bf 27},  598  (1983).

\bibitem{Halsey_XY}
T.~C. Halsey, J. Phys. C {\bf 18},  2437  (1985).

\bibitem{Korshunov_XY}
S.~E. Korshunov, J. Stat. Phys. {\bf 43},  17  (1986).

\bibitem{Granato_86}
E. Granato and {J. M. Kosterlitz}, Phys. Rev. B {\bf 33},  4767  (1986).

\bibitem{Granato_91}
E. Granato, {J. M. Kosterlitz}, J. Lee, and {M. P. Nightingale}, Phys. Rev.
   Lett. {\bf 66},  1090  (1991).

\bibitem{Li_Ciepak}
M.~S. Li and M. Cieplak, Phys. Rev. B {\bf 50},  955  (1994).

\end{thebibliography}
%\bibliographystyle{prsty}
%
%
%

\end{document}